\documentclass[showpacs,showkeys]{revtex4}

\usepackage{graphicx}
\usepackage{amsmath}
\usepackage{amssymb}
\usepackage{array}
\usepackage{color}
\newtheorem{thm}{}
\newcolumntype{C}{>{$}c<{$}}

\begin{document}

\title{Hidden nonassociative structure in supersymmetric quantum mechanics}

\author{Vladimir Dzhunushaliev
}
\affiliation{Department of Physics and Microelectronic
Engineering, Kyrgyz-Russian Slavic University, Bishkek, Kievskaya Str.
44, 720021, Kyrgyz Republic \\ 
and \\
Institute of Physics of National Academy of Science
Kyrgyz Republic, 265 a, Chui Street, Bishkek, 720071,  Kyrgyz Republic}
\email[Email: ]{vdzhunus@krsu.edu.kg}

\date{\today}

\begin{abstract}
It is shown that the Hamilton equations in supersymmetric quantum mechanics can be presented in nonassociative form, where the Hamiltonian is decomposed into two nonassociative factors.
\end{abstract}


\maketitle

\section{Introduction}

There are several equivalent formulations of quantum mechanics, including operator, matrix, and path integral descriptions. In this paper we will show that an equivalent formulation is possible for supersymmetric quantum mechanics, using a nonassociative commutator.

This effort relates to the fundamental question on how many descriptions of quantum mechanics there exist. In the 1930s, Pascual Jordan proposed a product
\begin{equation}
  x \cdot y = \frac{1}{2} ( xy + yx),
\label{0-40}
\end{equation}
such that $x \cdot y$ form an observable, where both $x$ and $y$ are observables themselves. This new product $( \cdot )$ forms a Jordan algebra of observables. In general, the classification of Jordan algebras yields all possible descriptions quantum mechanics, and how they differ.

The famous theorem by Jordan, von Neuman, and Wigner \cite{jordan} classifies the finite dimensional Jordan algebras: There exist only two, a special Jordan algebra and a 27-dimensional exceptional one.

Since physically relevant observables in quantum theory are infinite dimensional, the theorem by Jordan, von Neuman, and Wigner is not applicable there. Later, Zelmanov \cite{zelmanov} proved that there is no infinite dimensional simple exceptional Jordan algebra. While Zelmanov's theorem appears to destroy Jordan's original hope of constructing a new type of quantum mechanics, in this paper we propose a way to avoid the theorem's problematic consequences.

In our approach we consider an associative subalgebra 
$\mathbb G$ (algebra of observables) of a nonassociative algebra $\mathbb A$. The Hamilton equations then allow to formulate quantum dynamics for operators $L \in \mathbb G$, using nonassociative elements 
$h_1, h_2 \in \mathbb A$ ($h_1, h_2 \notin \mathbb G$).

As it is shown in Ref.\cite{Dzhunushaliev:2007cx}, the $h_{1,2}$ are unobservables, as they are modeled by nonassociative operators. There is an important distinction between such unobservables and the traditional understanding of "hidden variables" in quantum mechanics: In a way, the nonassociative variables $h_{1,2}$ could be understood as hidden parameters, as they are not visible to the experiment; however, while a hidden variables theory claims that its inner parameters \textcolor{red}{\emph{could be measured in principle}} , these nonassociative constituents $h_{1,2}$ \textcolor{red}{\emph{cannot be measured in principle.}}

\section{A nonassociative supersymmetric quantum mechanics}

The Hamilton equation in quantum mechanics,
\begin{equation}
  \frac{d L}{dt} = \frac{\partial L}{\partial t} + i \left[ H, L \right],
\label{1-30}
\end{equation}
is the dynamic equation of an operator $L$, where $H$ is the Hamiltonian. In Ref. \cite{okubo1995} the question is considered: What is the most general nonassociative algebra $\mathbb A$ which is compatible with Eq. \eqref{1-30} ?

The consistency condition 
\begin{equation} 
  \frac{d (xy)}{dt} = \frac{d x}{dt} y + x \frac{d y}{dt} , 
\label{1-40} 
\end{equation} 
requires validity of:
\begin{equation} 
  \left[ H, xy \right] = x \left[ H, y \right] + \left[ H, x \right] y. 
\label{1-50} 
\end{equation} 
But validity of Eq. \eqref{1-50} is not obvious for a general algebra. In particular, the following theorem must be considered:
\begin{thm}{\textbf{Theorem}} \label{theorem}
	(Myung \cite{myung}) 	
	To have a necessary and sufficient condition for 
	\begin{equation} 
	  \left[ z, xy \right] = x \left[ z, y \right] + \left[ z, x \right] y. 
	\label{1-60} 
	\end{equation} 
	for any $x,y,z \in \mathbb A$, it is required that $\mathbb A$ is flexible and Lie-admissible, i.e. 
	\begin{eqnarray} 
	 (x,y,z) &=& - (z,y,x), 
	\label{1-70}\\ 
	 \left[ \left[ x,y \right] z \right] + \left[ \left[ z,x \right], y \right] + 
	 \left[ \left[ y,z \right], x \right] &=& 0. 
	\label{1-80} 
	\end{eqnarray} 
\end{thm}
This theorem puts limitations onto allowable algebras $\mathbb A$, which can only be circumvented by violating the theorem's initial assumptions. Namely, Theorem \ref{theorem} requires that Eq. \eqref{1-60} is valid \emph{for any} $x,y,z \in \mathbb A$. We will relax this condition as follows.

Let $\mathbb{G} \subset \mathbb{A}$ be an associative subalgebra of the nonassociative algebra $\mathbb{A}$ . We then define a nonassociative commutator for $\mathbb G$ as:
\begin{equation}
	[g;h_1,h_2] := (g h_1) h_2 - h_1 (h_2 g),
\label{1-90}
\end{equation}
where $g \in \mathbb{G}$, $h_1, h_2 \in \mathbb A$.

For the special case $h_1, h_2 \in \mathbb G$, this reduces to the usual definition for the commutator:
\begin{equation}
	[g;h_1,h_2] = [g,H] = g H - H g,
\label{1-100}
\end{equation}
where $H = h_1 h_2$. For the opposite case, $h_{1,2} \notin \mathbb G $, the $h_{1,2}$ are now understood as nonassociative decomposition of the Hamiltonian, $H = h_1 h_2$.

Using the nonassociative commmutator \eqref{1-90}, we can propose a new description of quantum mechanics, with the modified Hamilton equation:
\begin{equation} 
  \frac{d L}{dt} = \frac{\partial L}{\partial t} - 
  i \left[ L; h_1, h_2 \right] .
\label{1-110} 
\end{equation} 
Now the question needs to be answered: Are the two descriptions of quantum mechanics with Hamilton equations \eqref{1-30} and \eqref{1-110} equivalent?

\subsection{Example 1. A nonassociative commutator for supersymmetric quantum mechanics}
\label{extension}

Let $\mathbb A = \mathbb {Q_M} \otimes \mathbb O$ where $\mathbb {Q_M}$ is from standard operator quantum mechanics, with operators $x$ and $p_i = - i \frac{\partial}{\partial x_i}$, and commutators:
\begin{equation}
	\left[ x_i, p_j \right] = i \delta_{ij}.
\label{1c-10}
\end{equation}
$\mathbb O$ is a split-octonion algebra (for details, see Appendix \ref{supersymmetry}). The associative subalgebra $\mathbb G$ consists of the following elements:
\begin{equation}
	L_0 u_0 + L_0^* u_0^* \in \mathbb G,
\label{1c-20}
\end{equation}
where $L_0, L_0^* \in \mathbb {Q_M}$. Let us to introduce operators \cite{Dzhunushaliev:2007vg} 
\begin{eqnarray}	
	Q &=& \sum \limits_{i=1}^3 \left( - p_j + i V_{,j} \right) u^*_j =
	\sum \limits_{i=1}^3 \mathcal D_j u^*_j ,
\label{1c-30}\\
	\bar Q &=& \sum \limits_{i=1}^3 \left( p_j + i V_{,j} \right) u_j =
	\sum \limits_{i=1}^3 \bar{\mathcal D}_j u_j ,
\label{1c-40}
\end{eqnarray}
with $V_{,j} = \frac{\partial V}{\partial x_j}$. The operators $Q, \bar Q$ are a nonassociative generalization of supersymmetric operator quantum mechanics (for details see Appendix \ref{split}).

We now define operators 
\begin{equation}
	h_1 := h_2 := Q + \bar Q , 
\label{1c-50}
\end{equation}
and calculate the nonassociative commutator from Eq. \eqref{1-90} as:
\begin{equation}
	\left[ L; \left( Q + \bar Q \right), \left( Q + \bar Q \right) \right] = 
	\left[ L; Q, \bar Q \right] + \left[ L; \bar Q, Q \right] = 
	\left[ \mathcal D_j \bar{\mathcal D}_j, L_0^* \right] u_0^* + 
	\left[ \bar{\mathcal D}_j \mathcal D_j, L_0 \right] u_0 ,
\label{1c-60}
\end{equation}
because $\left[ L; Q , Q \right] = \left[ L; \bar Q , \bar Q \right] = 0$. If 
\begin{eqnarray}
	L &=& A B ,
\label{1c-70}\\
	A &=& A_0 u_0 + A_0 u_0^* ,
\label{1c-80}\\
	B &=& B_0 u_0 + B_0 u_0^* ,
\label{1c-90}
\end{eqnarray}
then 
\begin{eqnarray}
	L_0 &=& A_0 B_0 ,
\label{1c-100}\\
	L_0^* &=& A_0^* B_0^* ,
\label{1c-110}
\end{eqnarray}
and from \eqref{1c-60} we see that:
\begin{equation}
	\left[ AB; \left( Q + \bar Q \right), \left( Q + \bar Q \right) \right] = 
	A \left[ B; \left( Q + \bar Q \right), \left( Q + \bar Q \right) \right] + 
	\left[ A; \left( Q + \bar Q \right), \left( Q + \bar Q \right) \right] B .
\label{1c-120}
\end{equation}
This satisfies the consistency condition \eqref{1-90} as required.

The Hamiltonian $H = \frac{1}{2} h_1 h_2$ can be decomposed in following way 
\begin{equation}
	H = \frac{1}{2} \left( Q + \bar Q \right)^2 = 
	\frac{1}{2} \left\{ \bar Q, Q \right\} = 
	\frac{1}{2} \left[
		{\hat p}^2 + \sum \limits_{j=1}^3 \left( V_{,j} \right)^2
	\right] + \frac{1}{2} \left( - i e_7 \right) \sum \limits_{j=1}^3 V_{,jj} ,
\label{1c-130}
\end{equation}
where $- i e_7 = u_0^* - u_0$ and $1 = u_0^* + u_0$ according to \eqref{app1-30}. 

This yields:
\begin{equation}
	[L, H] = \left[  
		L; \left( Q + \bar Q \right), \left( Q + \bar Q \right) 
	\right].
\label{1c-140}
\end{equation}
which means that the nonassociative commutators from \eqref{1c-60} and \eqref{1c-140} are equivalent. Therefore, we conclude that both descriptions of supersymmetrical quantum mechanics based on Hamilton equations \eqref{1c-60} and \eqref{1c-140} are equivalent.

\subsection{Example 2. Nonassociative commutator for quaternions}
\label{qm_quaternions}

Let $\mathbb G = \mathbb Q$ and $\mathbb A = \mathbb O$ be quaternions and octonions respectively (for definition and multiplication tables see Appendix \ref{app1} ).

In this case, we have $g = i_n$, $h_1 = i_4$, $h_2 = i_{m+4}$, with $m,n = 1,2,3$. The consistency condition \eqref{1-90} is satisfied:
\begin{equation}
	\left[i_k i_l; i_4,i_{m+4} \right] = 
	i_k \left[i_l; i_4, i_{m+4} \right] + 
	\left[i_k; i_4, i_{m+4} \right] i_l, \quad m,k,l = 1,2,3. 
\label{1a-10}
\end{equation}

\subsection{Example 3. A nonassociative commutator for biquaternions}
\label{extended}

A construction similar to the section \ref{qm_quaternions} above can be done for the biquaternions as well. In this case, the consistency condition \eqref{1-90} can be proven by direct calculation from Table~\ref{sedenions} :
\begin{equation}
	\left[ ab;	i_4, \epsilon_{m+4} \right] = 
	a \left[ b;	i_4, \epsilon_{m+4} \right] + 
	\left[ a;	i_4, \epsilon_{m+4} \right] b ,
\label{1b-10}
\end{equation}
for any $a \in \left\{ i_k, \epsilon_l \right\}$ and $b \in \left\{ i_k, \epsilon_l \right\}$ with $k,l \in \left\{ 1,2,3 \right\} $.

\section{Conclusions}

We have shown in example \ref{extension} that there exists a nonassociative structure in supersymmetric quantum mechanics, in the sense that we can: (a) rewrite the Hamiltion equation \eqref{1-30} using a nonassociative commutator; and (b) the Hamiltonian can be decomposed as the product of nonassociative factors. We conclude that this forms a possible, non-standard description of supersymmetrical quantum mechanics. It is interesting to note that in Ref.\cite{Kuznetsova:2005cd} is found another hidden non-associative structure which manifests that in a N=8 supersymmetric quantum mechanics (the (1,8,7) model) the octonionic structure constants $C_{ijk}$ enter as coupling constants in the invariant action. The model itself is associative but the non-associativity is hidden in its coupling constants.

In order to distinguish this from other formulations of quantum mechanics, it is important to investigate the following question: Does there exist a nonassociative algebra $\mathbb A$ with associative subalgebra $\mathbb G$, such that the Hamilton equations \eqref{1-30} and \eqref{1-110} are \emph{not} equivalent?

The exitence of a nonassociative commutator tells us that a hidden structure in supersymmetric quantum mechanics is possible, formed by operators $h_{1,2}$ which are unobservable due to their nonassociativity \cite{Dzhunushaliev:2007vg} . It is necessary to recall that this hidden structure is not a "hidden variables" theory, where the presumed inner parameters \textcolor{red}{\emph{could be measured in principle}}. Instead, the nonassociative constituents $h_{1,2}$ presented here \textcolor{red}{\emph{cannot be measured in principle}}.

From the author's point of view, a similar hidden structure may relate to the confinement problem in quantum chromodynamics. The problem there arises from the fact that, for strongly interacting quantum fields, we do not know a corresponding algebra of field operators. We believe that the construction of such algebra may be connected with an unobservable, nonassociative structure.

\section{Acknowledgments} 

I am grateful to the Alexander von Humboldt Foundation for the support, and to J. K\"oplinger for fruitful discussion.

\appendix

\section{Supersymmetric quantum mechanics}
\label{supersymmetry}

In this section we follow Ref.~\cite{Haymaker:1985us}. A one-dimensional quantum mechanical Hamiltonian 
\begin{equation}	
	H = \left(
	\begin{array}{cc}
		H_-	&	0	\\
		0					&	H_+
	\end{array}
	\right)
\label{app0-10}
\end{equation}
is said to be supersymmetric \cite{Witten:1981nf} \cite{Cooper:1982dm} if the corresponding potentials $V_\pm (x)$ are related according to 
\begin{equation}	
	V_{\pm} = \frac{{U'}^2}{8} \mp \frac{U''}{4}.
\label{app0-20}
\end{equation}
The demonstration that $H$ is supersymmetric hinges on the existence of the generators of supersymmetry $Q, \bar Q$ which together with $H$ satisfy the commutation and anticommutation relations
\begin{eqnarray}	
	\left[ Q, H \right] &=& \left[ \bar Q, H \right] = 0 ,
\label{app0-30}\\
	\left\{ \bar Q, \bar Q \right\} &=& \left\{ Q, Q \right\} = 0 ,
\label{app0-40}\\
	\left\{ \bar Q, Q \right\} &=& 2 H
\label{app0-50}
\end{eqnarray}
here
\begin{eqnarray}	
	Q &=& \left( p - i \frac{U'}{2} \right) \sigma^+ , \quad 
	\sigma^+ = \left(
	\begin{array}{cc}
		0	&	1	\\
		0	&	0
	\end{array}
	\right),
\label{app0-60}\\
	\bar Q &=& \left( p + i \frac{U'}{2} \right) \sigma^- , \quad 
	\sigma^- = \left(
	\begin{array}{cc}
		0	&	0	\\
		1	&	0
	\end{array}
	\right) 
\label{app0-70}
\end{eqnarray}
where $p = -i \frac{\partial}{\partial x}$. Because of the relations $\left\{ \sigma^-, \sigma^+ \right\} = 1$ and
$\left[ \sigma^+, \sigma^- \right] = \sigma_z$, it is easily verified that Eqs.~\eqref{app0-30}-\eqref{app0-50} are satisfied, and that
\begin{equation}	
	H = \frac{1}{2} \left(
		Q \bar Q + \bar Q Q
	\right) =
	\frac{1}{2} \left(
		p^2 + \frac{{U'}^2}{4}
	\right) \mathbb I +
	\frac{U''}{4} \sigma_z ,
\label{app0-100}
\end{equation}
where $\mathbb I$ is the identity matrix and $\sigma_z$ is the Pauli matrix.

\section{The split-octonion algebra}
\label{split}

In this section we follow Ref.~\cite{Gunaydin:1973rs}. A composition algebra is defined as an algebra $A$ with identity element and a nondegenerate quadratic form $Q$ over $A$, such that $Q$ permits the composition
\begin{equation}
	Q(xy) = Q(x) Q(y), \; x,y \in A.
\label{app1-10}
\end{equation}
According to the Hurwitz theorem, only four different composition algebras exist over the real or complex number fields. These are the real numbers $\mathbb{R}$ of dimension 1, complex numbers $\mathbb{C}$ of dimension 2, quaternions $\mathbb{H}$ of dimension 4, and octonions $\mathbb{O}$ of dimension 8. Of these algebras, the quaternions $\mathbb{H}$ are not commutative and the octonions $\mathbb{O}$ are neither commutative nor associative. A composition algebra is said to be a division algebra if the quadratic form $Q$ has the following property
\begin{equation}
	\text{if } Q(x) = 0 \text{ implies that } x=0.
\label{app1-20}
\end{equation}
Otherwise, the algebra is called \textit{split}.
\par
A basis for a real octonion $\mathbb{O}$ contains eight elements, including identity:
\begin{equation}
	1, e_A, \; A=1, \cdots , 7, \; \text{ where } e^2_A=-1.
\label{app1-21}
\end{equation}
The elements $e_A$ satisfy the following multiplication table: 
\begin{equation}
	e_A e_B = a_{ABC} e_C - \delta_{AB}
\label{app1-22}
\end{equation}
where $a_{ABC}$ is totally antisymmetric and
\begin{equation}
	a_{ABC} = +1 \text{ for } ABC = 123, 516, 624, 435, 471, 673, 572.
\label{app1-23}
\end{equation}
For the split-octonion algebra we choose the following basis: 
\begin{equation}
\begin{split}
	u_i = &\frac{1}{2} \left( e_i + i e_{i+3} \right),	\quad
	u_i^* = \frac{1}{2} \left( e_i - i e_{i+3} \right)	, \quad
	i=1,2,3 ;
\\
	u_0 = &\frac{1}{2} \left( 1 + i e_7 \right),	\quad
	u_0^* = \frac{1}{2} \left( 1 - i e_7 \right)	.
\label{app1-30}
\end{split}
\end{equation}
These basis elements satisfy the multiplication table
\begin{align}
	u_i u_j 	&= \epsilon_{ijk} u^*_k,& u^*_i u^*_j &= \epsilon_{ijk} u_k, &i,j,k 		 &= 1,2,3 	&&
\label{app1-40} \\
	u_i u^*_j &= - \delta_{ij} u_0, 	&u^*_i u_j 		&= - \delta_{ij} u^*_0,&					 &					&&
\label{app1-50}\\
	u_i u_0 	&= 0, 									&u_i u^*_0 		&= u_i	,							 &u^*_i u_0 &= u_i^*	,		&u_i^* u_0^* &=0,
\label{app1-60}\\
	u_0 u_i 	&= u_i, 								&u^*_0 u_i 		&= 0	,	  						 &u_0 u^*_i &= 0	,				&u_0^* u_i^* &=u_i^*,
\label{app1-70}\\
	u_0^2 	&= u_0,   								&{u^*_0}^2 		&= u^*_0	,	  						 &u_0 u^*_0 &= u_0^* u_0 = 0	.				&&
\label{app1-80}
\end{align}
The split octonion algebra contains divisors of zero and hence is not a division algebra. 

\section{Sedenions}
\label{app1}

Sedenions after \cite{Carmody} (not to be confused with sedenions from Cayley-Dickson construction) form an algebra with nonassociative but alternative multiplication, and a multiplicative modulus. It consists of one real axis (to basis $1$), eight imaginary axes (to bases $i_n$ with $i_n^2 = -1, n=0, \ldots , 7$), and seven axes to non-real roots of $+1$, i.e., bases $\epsilon_n$ with $\epsilon_n^2 = +1, n=1, \ldots , 7$. The multiplication table is given in Table~\ref{sedenions}. These sedenions are isomorphic to octonions with complex coefficients, $\mathbb{C}\times\mathbb{O}$, and contain following important subalgebras:
\begin{itemize}
	\item the associative quaternion subalgebra $\mathbb Q$ with $\left\{ 1, i_n \right\}$, $n = 1,2,3$;
	\item the associative biquaternion subalgebra $\mathbb B$ with 
	$\left\{ 1 , i_0 , i_n , \epsilon_n \right\}$, $n = 1,2,3$;
	\item the nonassociative octonion subalgebra $\mathbb O$ with $\left\{ 1, i_n \right\}$, $n = 1, \ldots 7$,
	\item the nonassociative split-octonion subalgebra with $\left\{ 1, i_n, \epsilon_4, \epsilon_{(n+4)} \right\}$, $n = 1,2,3$.
\end{itemize}
\begin{table}[h]
\begin{tabular}{|
C|| C|| C| C| C| C| C| C| C|| C|| C| C| C| C| C| C| C| C|} \hline
			&	  1  					&  i_1  				&  i_2  				&  i_3   				& i_4  			&  i_5  				&  i_6   				&	 i_7  				&	 i_0   				
			&  \epsilon_1  	&  \epsilon_2  	&  \epsilon_3   &  \epsilon_4  	&  \epsilon_5  	& 							 \epsilon_6   	&	 \epsilon_7  	  
\\ 
\hline \hline
1			&		  1  				&  i_1  					&  i_2  				&  i_3   				&		  i_4  			&  i_5  			&  i_6   				&	 i_7  					&	 i_0   				
			&  \epsilon_1  	&  \epsilon_2  		&  \epsilon_3   &  \epsilon_4  	&  \epsilon_5  	& 							 \epsilon_6   	&	 \epsilon_7  	  
\\ 
\hline \hline
i_1		&	 i_1  				&  -1  						&  i_3  				&  -i_2   			&		  i_5  			& -i_4 				&  -i_7   			&	 i_6 	&	 -\epsilon_1   					&  i_0  				&  \epsilon_3  
			&  -\epsilon_2  &  \epsilon_5  		&  -\epsilon_4  &	 -\epsilon_7  &	 \epsilon_6   
\\ 
\hline 
i_2		&	 i_2  				&  -i_3  					&  -1  					&  i_1   				&		  i_6  			
			& i_7 					&  -i_4   				&	 -i_5 				&	 -\epsilon_2  &  -\epsilon_3  							&  i_0  				&  \epsilon_1 		&  \epsilon_6  	&  \epsilon_7  	&	 -\epsilon_4  
			&	 -\epsilon_5   
\\ 
\hline 
i_3 	&	 i_3  				&  i_2  					&  -i_1 				&  -1  					&  i_7   				
			&	-i_6 & i_5 		&  -i_4   				&	 -\epsilon_3 	&	 \epsilon_2  	&  -\epsilon_1  
			&  i_0  				&  \epsilon_7 		&  -\epsilon_6  &  \epsilon_5  	&	 -\epsilon_4    
\\ 
\hline 
 i_4 	&	 i_4  				&  -i_5  					&  -i_6 				&  -i_7  					
			&  -1   				&	 i_1 						& i_2 					&  i_3   		
			&	 -\epsilon_4 	 
			&	 -\epsilon_5  &  -\epsilon_6  	&  -\epsilon_7 	&  i_0  
			&  \epsilon_1 	&  \epsilon_2  		&  \epsilon_3  		
\\ 
\hline 
 i_5 	&	 i_5  				&  i_4  					&  -i_7 				&  i_6  					
			&  -i_1   			&	 -1 						& -i_3 					&  i_2   		
			&	 -\epsilon_5 	 
			&	 \epsilon_4  	&  -\epsilon_7  	&  \epsilon_6 	&  -\epsilon_1  
			&  i_0 					&  -\epsilon_3  	&  \epsilon_2  		
\\ 
\hline 
 i_6 	&	 i_6  				&  i_7  					&  i_4 				&  -i_5  					
			&  -i_2   			&	 i_3 						& -1 					&  -i_1   		
			&	 -\epsilon_6 	 
			&	 \epsilon_7  	&  \epsilon_4  		&  -\epsilon_5 &  -\epsilon_2  
			&  \epsilon_3 	&  i_0  					&  -\epsilon_1  		
\\ 
\hline 
 i_7 	&	 i_7  				&  -i_6  					&  i_5 				&  i_4  					
			&  -i_3   			&	 -i_2 					& i_1 				&  -1   		
			&	 -\epsilon_7 	 
			&	 -\epsilon_6  &  \epsilon_5  		&  \epsilon_4 &  -\epsilon_3  
			&  -\epsilon_2 	&  \epsilon_1  		&  i_0  		
\\ 
\hline \hline
 i_0 	&	 i_0  				&  -\epsilon_1  	&  -\epsilon_2 &  -\epsilon_3  					
			&  -\epsilon_4  &	 -\epsilon_5 		& -\epsilon_6  &  -\epsilon_7   		
			&	 -1 	 
			&	 i_1  				&  i_2  					&  i_3 					&  i_4  
			&  i_5 					&  i_6  					&  i_7  		
\\ 
\hline \hline
 \epsilon_1 	
			&	 \epsilon_1  	&  i_0  					&  \epsilon_3 	&  -\epsilon_2  					
			&  \epsilon_5  	&	 -\epsilon_4 		& -\epsilon_7 	&  \epsilon_6   		
			&	 i_1 	 
			&	 1  					&  -i_3  					&  i_2 					&  -i_5  
			&  i_4 					&  i_7  					&  -i_6  		
\\ 
\hline 
 \epsilon_2 	
			&	 \epsilon_2  	&  -\epsilon_3  	&  i_0 					&  \epsilon_1  		
			&  \epsilon_6  	&	 \epsilon_7 		& -\epsilon_4 	&  -\epsilon_5   		
			&	 i_2 	 
			&	 i_3  				&  1  						&  -i_1 				&  -i_6  
			&  -i_7 				&  i_4  					&  i_5  		
\\ 
\hline 
 \epsilon_3 	
			&	 \epsilon_3  	&  \epsilon_2  		&  -\epsilon_1 	&  i_0  		
			&  \epsilon_7  	&	 -\epsilon_6 		& \epsilon_5 		&  -\epsilon_4   		
			&	 i_3 	 
			&	 -i_2  				&  i_1  					&  1 						&  -i_7  
			&  i_6 					&  -i_5  					&  i_4  		
\\ 
\hline 
 \epsilon_4 	
			&	 \epsilon_4  	&  -\epsilon_5  	&  -\epsilon_6 	&  -\epsilon_7  		
			&  i_0  				&	 \epsilon_1 		& \epsilon_2 		&  \epsilon_3   		
			&	 i_4 	 
			&	 i_5  				&  i_6  					&  i_7 					&  1  
			&  -i_1 				&  -i_2  					&  -i_3  		
\\ 
\hline 
 \epsilon_5 	
			&	 \epsilon_5  	&  \epsilon_4  		&  -\epsilon_7 	&  \epsilon_6  		
			&  -\epsilon_1  &	 i_0 						& -\epsilon_3 	&  \epsilon_2   		
			&	 i_5 	 
			&	 -i_4  				&  i_7  					&  -i_6 				&  i_1  
			&  1 						&  i_3  					&  -i_2  		
\\ 
\hline 
 \epsilon_6 	
			&	 \epsilon_6  	&  \epsilon_7  		&  \epsilon_4 	&  -\epsilon_5  		
			&  -\epsilon_2  &	 \epsilon_3 		& i_0 					&  -\epsilon_1   		
			&	 i_6 	 
			&	 -i_7  				&  -i_4  					&  i_5 					&  i_2  
			&  -i_3 				&  1  						&  i_1  		
\\ 
\hline 
 \epsilon_7 	
			&	 \epsilon_7  	&  -\epsilon_6  	&  \epsilon_5 	&  \epsilon_4  		
			&  -\epsilon_3  &	 -\epsilon_2 		& \epsilon_1 		&  i_0   		
			&	 i_7 	 
			&	 i_6  				&  -i_5  					&  -i_4 				&  i_3  
			&  i_2 					&  -i_1  					&  1  		
\\ 
\hline 
\end{tabular}
\caption{A multiplication table for sedenions.}
\label{sedenions}
\end{table}

\end{document}